\documentclass[fleqn,twoside]{article}
\usepackage{espcrc2}
\usepackage{amssymb}
\usepackage[figuresright]{rotating}

\newcommand{\AmS}{{\protect\the\textfont2
  A\kern-.1667em\lower.5ex\hbox{M}\kern-.125emS}}

\newcommand{\mon}{{\mathrm{mon}}}

\newcommand{\Z}{{Z \!\!\! Z}}
\newcommand{\be}{\begin{equation}}
\newcommand{\ee}{\end{equation}}
\newcommand{\beqn}{\begin{eqnarray}}
\newcommand{\eeqn}{\end{eqnarray}}
\newcommand{\eq}[1]{(\ref{#1})}

\title{
\vspace{-25mm}
\rightline{\small KANAZAWA-03-23~~~~~}
\rightline{\small ITEP-LAT-2003-18~~~~~}
\vspace{10mm}
Energy-entropy study of projected space--like monopoles \newline
in finite--T quenched SU(2) 
QCD\thanks{Presented by T.~S. at Lattice'03.}
}

\author{Tsuneo Suzuki\address{Institute for Theoretical Physics, Kanazawa University,
Kanazawa 920-1192, Japan}\thanks{T.S. is partially
supported by JSPS Grant-in-Aid for Scientific Research on Priority Areas
No.13135210 and (B) No.15340073. This work is also supported by the
Supercomputer Project of the Institute of Physical and Chemical
Research (RIKEN).  A part of our numerical simulations have been done
using NEC SX-5 at Research Center for Nuclear Physics (RCNP) of Osaka
University.},
M. N. Chernodub${}^{\mathrm{a,}}$\address{ITEP, B.Cheremushkinskaya 25, Moscow,
117259, Russia}\thanks{M.N.Ch. is supported by JSPS Fellowship P01023.}
and Katsuya~Ishiguro${}^{\mathrm{a}}$}
\begin{document}

\begin{abstract}
Properties of space--like monopoles projected on the 3D space 
in finite temperature quenched SU(2) QCD are studied.
The monopole energy is derived from the effective action of the monopoles which is determined by an 
inverse Monte-Carlo method. Then the entropy is fixed with the help of the monopole--loop distribution.
\end{abstract}

\maketitle

\section{INTRODUCTION}

The dual superconductor mechanism~\cite{DualSuperconductor} of color confinement is confirmed in many
numerical simulations in the Maximal Abelian (MA) gauge~\cite{kronfeld} (for a review, see
Ref.~\cite{Reviews}). This mechanism is based on the existence of the Abelian monopoles which are
condensed in the confinement phase of QCD~\cite{shiba:condensation,MonopoleCondensation}. The monopole
condensate corresponds to the so-called percolating (infrared)
cluster~\cite{ivanenko,ref:kitahara} of the monopole trajectories.  The tension of the
confining string gets a dominant contribution from the IR cluster~\cite{ref:kitahara} while the
finite-sized (ultraviolet) clusters do not play any role in confinement. Various properties of the UV
and IR monopole clusters were studied previously in
Refs.~\cite{ref:kitahara,ref:clusters,IshiguroSuzuki}.

In the high temperature phase the monopoles become static. In this phase the IR cluster 
disappears~\cite{ivanenko,ref:kitahara} and, consequently, the confinement of the static quarks is 
absent because only spatial components of the IR monopole cluster are relevant for the 
confinement. We investigate the action, the length distribution and the entropy of spatial 
components of the infrared monopole clusters following Ref.~\cite{IshiguroSuzuki}.

\section{MODEL}

We use the Wilson action, $S(U) = - \frac{\beta}{2}\, {\mathrm{Tr}} U_P$, to generate
1000-3000 configurations of the SU(2) gauge field, $U$, for $\beta=2.3 \sim 2.6$
on $48^3\times N_t$, $N_t=6,8,12,16$ lattices. Performing the MA gauge fixing, 
we locate Abelian monopoles in the gluonic fields configurations in
a standard way. The phase of the diagonal
component of the SU(2) gauge field gives the Abelian gauge field, $\theta_\mu(s) = \arg\,
U^{11}_{\mu}(s)$. Then we construct the Abelian field strength,
$\theta_{\mu\nu}(s)\in(-4\pi,4\pi)$, which is decomposed into two parts, $\theta_{\mu\nu}(s)=
\bar{\theta}_{\mu\nu}(s) +2\pi m_{\mu\nu}(s)$. Here $\bar{\theta}_{\mu\nu}(s)\in [-\pi,\pi)$ 
and $m_{\mu\nu}(s)$ are the electromagnetic flux and the Dirac string coming through the plaquette 
$P_{\mu\nu}(s)$, respectively. The ends of the Dirac strings correspond to the
conserved elementary monopole currents~\cite{degrand}, $k_{\mu}(s) =1/2 \epsilon_{\mu\nu\rho\sigma}
\partial_{\nu}m_{\rho\sigma}(s+\hat{\mu}) \in \Z$, where $\partial$ is the forward lattice derivative. 
The elementary monopoles are defined on the fine lattice with the spacing $a$. To study the monopole charges 
at various scales $b=na$, we use the extended $n^3$ monopole construction~\cite{ivanenko},
\beqn
k_{\mu}^{(n)}(s) = \!\!\sum_{i,j,l=0}^{n-1}\!\!k_{\mu}(n s+(n-1)\hat{\mu}+i\hat{\nu}
     +j\hat{\rho}+l\hat{\sigma}).\nonumber
\eeqn
The spatially projected currents 
are:
\beqn
K_i^{(n)}(\vec s) = \sum\nolimits^{N_t-1}_{s_4=0}\, k_i^{(n)}(s,s_4)\,,\,\, i=1,2,3\,.
\eeqn

\section{MONOPOLE ACTION}
\label{sec:action}

The monopole action of the $3D$ projected IR monopole clusters can be defined using the inverse 
Monte--Carlo method~\cite{shiba:condensation}. The action is represented in a 
truncated form~\cite{shiba:condensation,chernodub} as a sum of the $n$--point ($n \ge 2$) operators 
$S_i$:
$$
S_{\mon}[k] = \sum\nolimits_i f_i S_i [k]\,,
\vspace{-2mm}
$$
where $f_i$ are the coupling constants. Following Ref.~\cite{IshiguroSuzuki}
we adopt only the two--point interactions in the monopole action ($i.e.$ interactions of the form $S_i 
\sim k_{i}(s) k_{j}(s')$). 
\vskip -6mm
\begin{figure}[h]
\centerline{\includegraphics[angle=0,scale=0.24,clip=true]{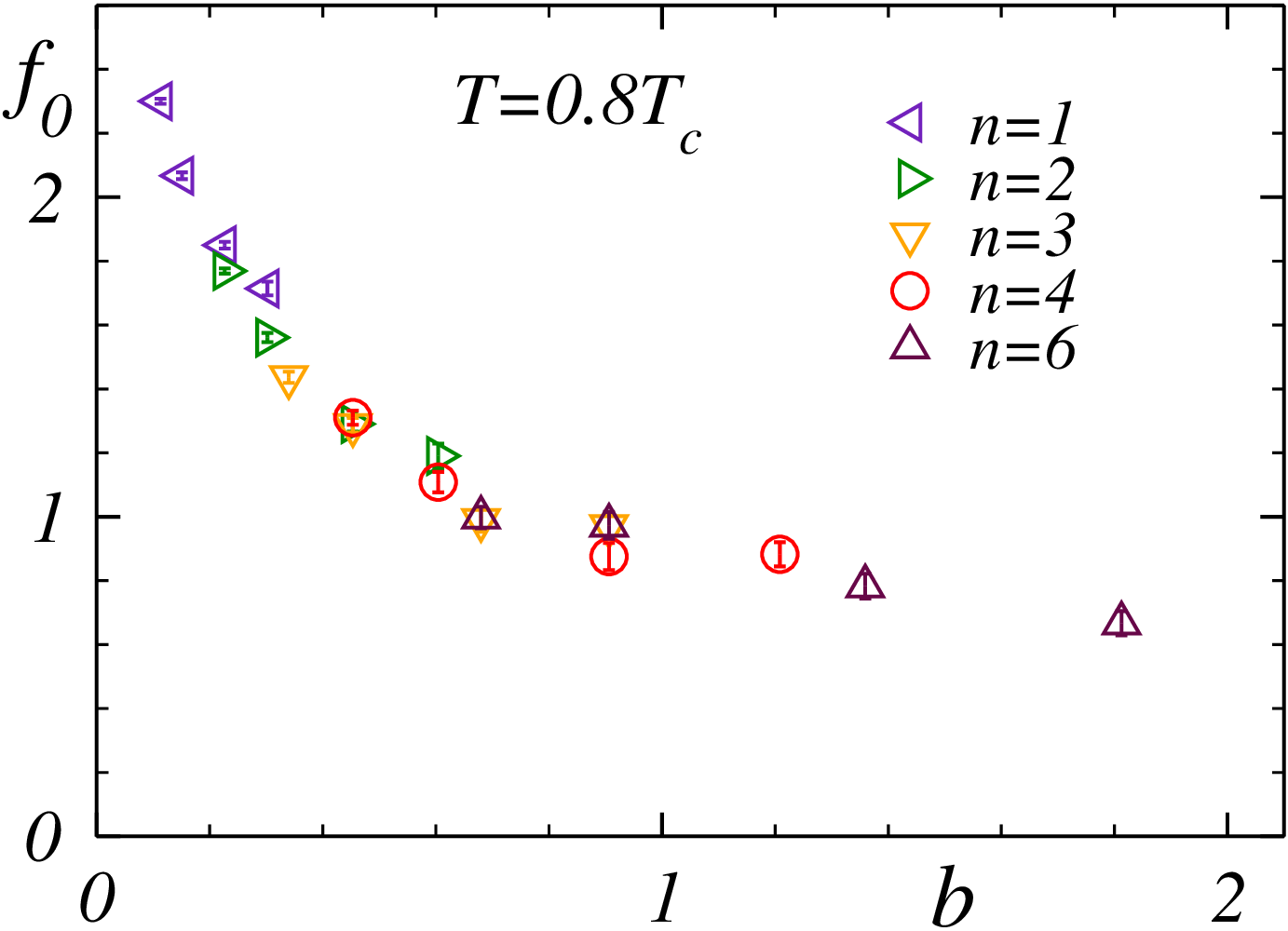}}
\vskip -8mm
\caption{The parameter $f_0$ $vs.$ $b$ at $T=0.8 T_c$.}
\label{fig:f0}
\end{figure}
\vskip -7mm

Similarly to the $4D$ case we find that the monopole action of the $3D$ currents is proportional with a
good accuracy to the length of the monopole loop, $S_{\mon} = f_0 L + const$. The
coupling $f_0$ shows a good scaling in agreement with Ref.~\cite{shiba:condensation}, as can be seen
from Fig.~\ref{fig:f0} where this coupling is shown as a function of the scale $b$ at $T=0.8 T_c$.

\section{LENGTH DISTRIBUTION}
\label{sec:length}

The length of the $4D$ IR monopole currents in the finite volume $V$ 
is distributed according to the Gaussian law~\cite{IshiguroSuzuki}:
\beqn
D^{IR}(L) \propto \exp\{ - \alpha(b,V) L^2 + \gamma(b,T) L\}\,.
\label{eq:IR:distr:two}
\eeqn
The length distribution function, $D(L)$, is proportional to the
weight with which the particular trajectory of the length $L$ contributes to 
the partition function. In Eq.~\eq{eq:IR:distr:two} we neglect a 
power-law prefactor which is essential for the distribution of the
ultraviolet clusters. 

Since the monopole density, $\rho_{IR} = L_{max}/ V$, is finite, the 
peak of the distribution~\eq{eq:IR:distr:two}, 
$L_{max} = \gamma(b,T) \slash 2 \, \alpha(b,V)$,
must be proportional to the volume of the system for large volumes.
Thus, in the thermodynamic limit we expect $\alpha(b,V) = A(b) \slash V$, 
where $A(b)$ is a certain function. Therefore
in the thermodynamic limit the parameter $\alpha$ vanishes.

Apart from the finite--volume effect, the distribution~\eq{eq:IR:distr:two} 
has contributions from the energy and the entropy. As  seen above, 
the action contribution is proportional to $e^{- f_0 L}$. The entropy 
contribution is proportional to $\mu^L$ (with $\mu>0$) for sufficiently 
large monopole lengths, $L$. Thus, the entropy factor, $\mu$, is
\beqn
\mu = \exp\{f_0 + \gamma\}\,.
\label{eq:mu}
\eeqn

\begin{figure}[thb]
\centerline{\includegraphics[angle=0,scale=0.24,clip=true]{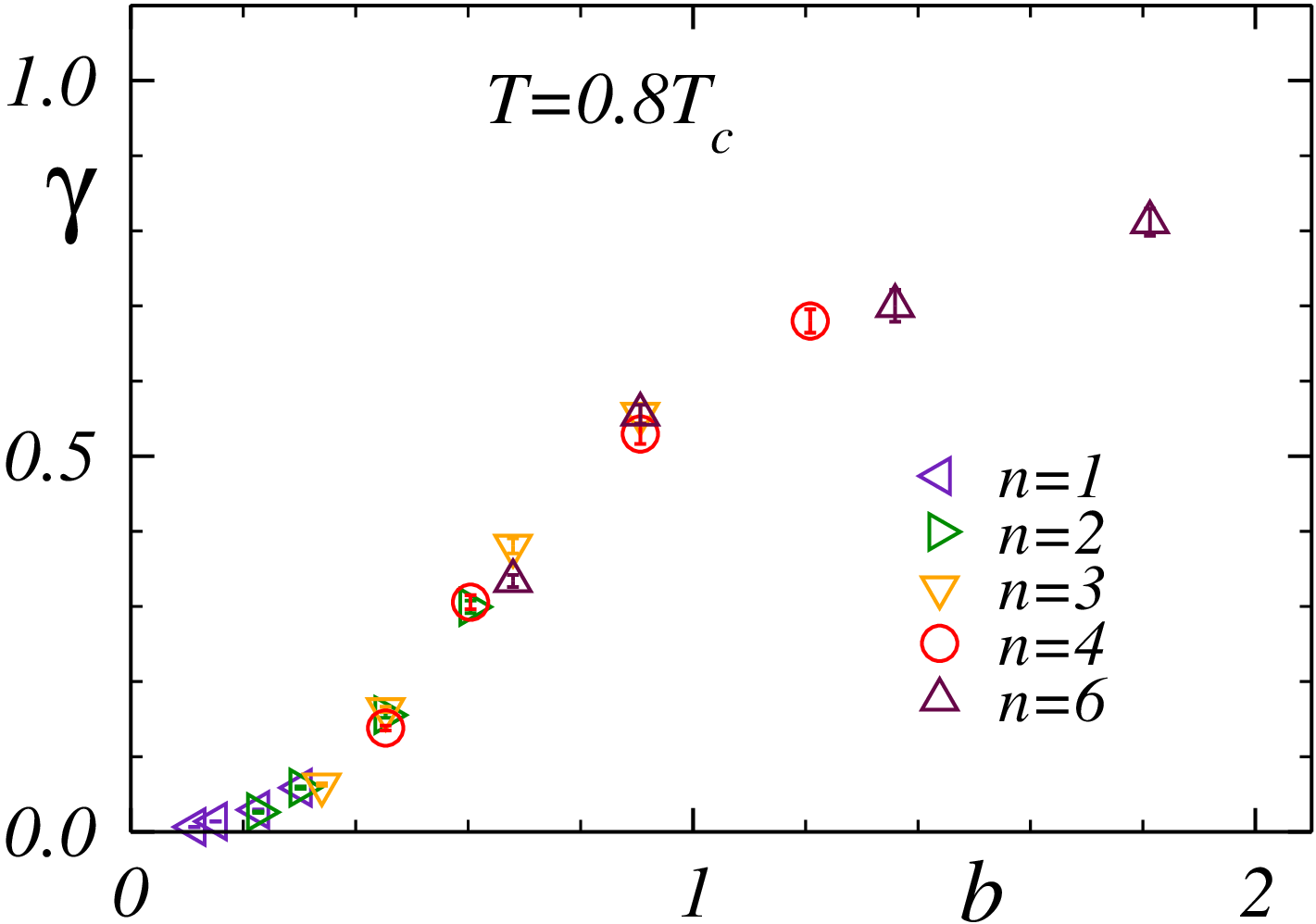}}
\vskip -6mm
\caption{The parameter $\gamma$ $vs.$ $b$ at $T=0.8 T_c$.}
\label{fig:gamma}
\vskip -8mm
\end{figure}

Since we are performing the simulations in a finite volume we 
fit numerically obtained distributions of the $3D$ projected currents
by the function~\eq{eq:IR:distr:two} 
and then use the bootstrap method to estimate the statistical errors. 
We find that the parameter $\gamma$ shows a good scaling with $b$.
An example at $T=0.8 T_c$ is shown in Fig.~\ref{fig:gamma}. 
In a small $b$--region
we find that $\gamma \propto b^\eta$ with $\eta \sim 3$ for low temperatures,
say, at
$T \sim 0.5 T_c$, whereas $\eta \sim 2$ for $T \to T_c$.
\vskip -6mm

\begin{figure}[thb]
\centerline{\includegraphics[angle=0,scale=0.22,clip=true]{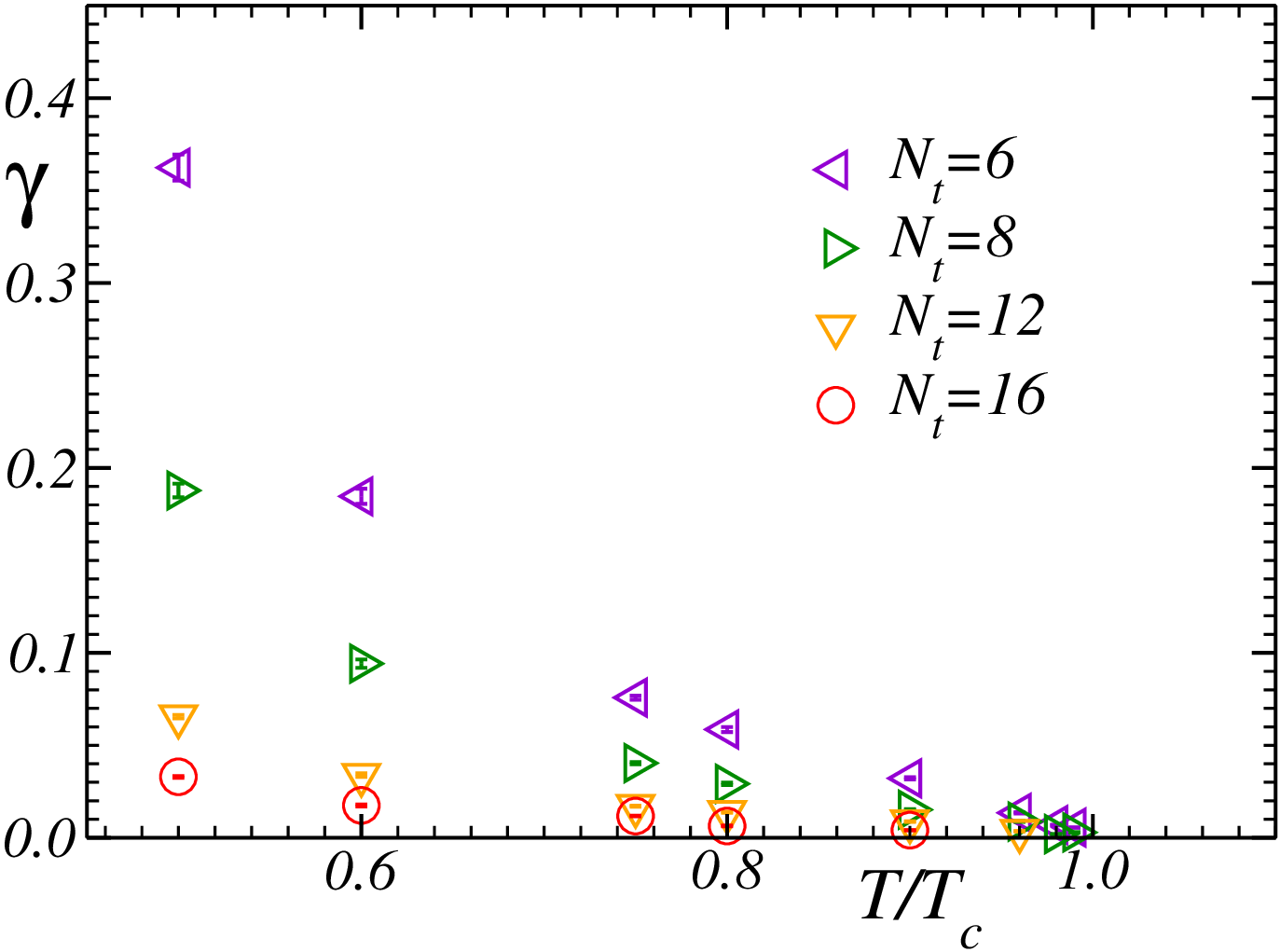}}
\vskip -8mm
\caption{The parameter $\gamma$ $vs.$ $T$ for $n=1$.}
\label{fig:gamma:T}
\vskip -6mm
\end{figure}

At the critical temperature, $T=T_c$, 
the $4D$ IR monopole cluster disappears and we expect
a similar behaviour for the $3D$ projected IR cluster. 
Thus the parameter $\gamma(b,T)$ 
should vanish at the critical point. This is indeed the case according to 
Fig.~\ref{fig:gamma:T} which  shows $\gamma$ for elementary ($n=1$) 
monopoles 
as a function of temperature for various temporal extensions of the lattice.
%

\section{MONOPOLE ENTROPY}

\begin{figure}[t]
\centerline{\includegraphics[angle=0,scale=0.22,clip=true]{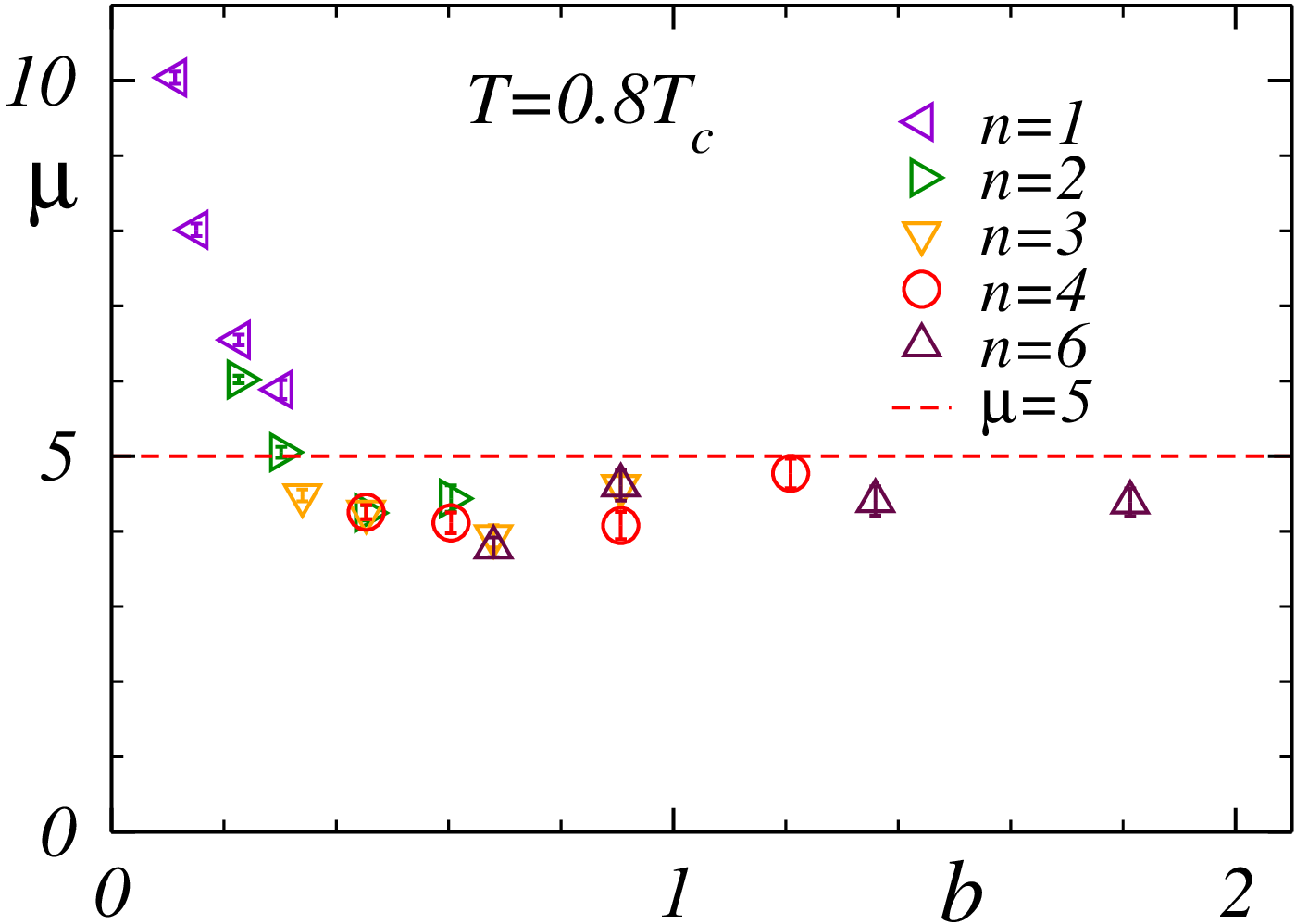}} 
\vskip 2mm
\centerline{\includegraphics[angle=0,scale=0.22,clip=true]{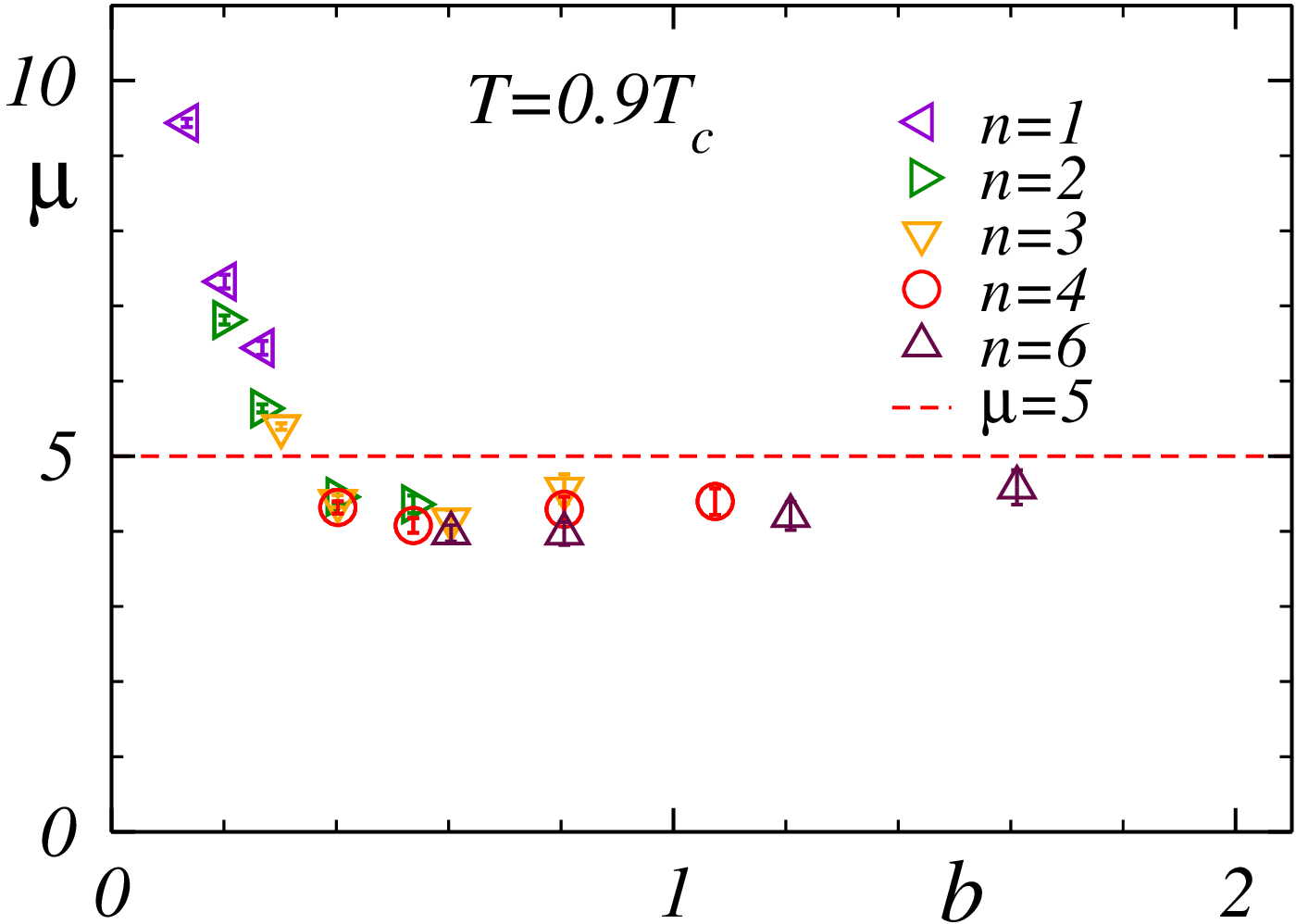}} 
\vskip 2mm
\centerline{\includegraphics[angle=0,scale=0.22,clip=true]{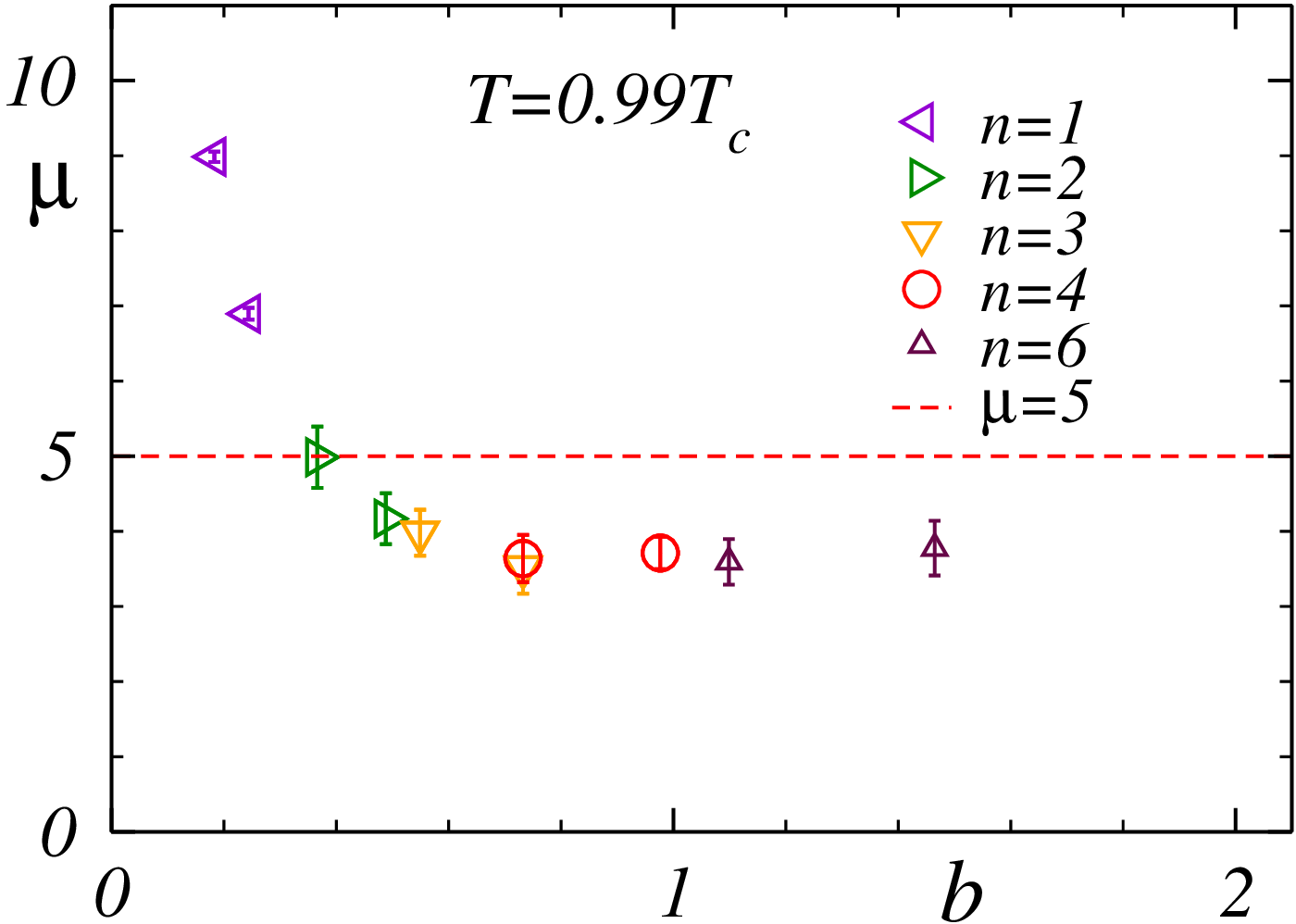}} 
\vskip -6mm 
\caption{The entropy factor $\mu$ $vs.$ the scale $b$ at temperatures $T/T_c=0.8,0.9$ and $0.99$.}
\label{fig:entropy} 
\vskip -6mm 
\end{figure}

We determine the entropy using Eq.~\eq{eq:mu}. The numerical results 
for the entropy factor $\mu(b,T)$ are shown in Fig.~\ref{fig:entropy} for 
various temperatures. 
If the monopoles are randomly walking on a $3D$
hypercubic lattice then we should get $\mu=5$ since five choices exist 
at each site for the monopole
current to go further. Note that $\mu>5$ for small values of $b$ 
because in this region the quadratic monopole
actions are not enough~\cite{chernodub}.

At large $b$ the entropy factor tends to a fixed value 
around 4, which
seems independent of temperature. Contrary to the 
$4D$ case~\cite{IshiguroSuzuki} 
the entropy $\mu$ for projected monopoles does not approach unity 
in the $b \to \infty$ limit. 
Thus we observe that the spatially projected monopole currents exhibit 
a motion close to the random walk at 
non--zero temperatures.

\end{document}